\documentclass[journal]{IEEEtran}
\usepackage{color}
\usepackage{cite}
\usepackage{amsmath,amsfonts}
\usepackage{algorithmic}
\usepackage{algorithm}
\usepackage{array}
\usepackage{textcomp}
\usepackage{stfloats}
\usepackage{url}
\usepackage{verbatim}
\usepackage{graphicx}
\usepackage{balance}
\usepackage{ulem}
\usepackage{float} 
\usepackage{subfigure}
\usepackage{booktabs}
\usepackage{setspace}
\usepackage{amssymb}
\usepackage[numbers,sort&compress]{natbib}
\usepackage{hyperref}
\usepackage{ulem}

\begin{document}
\graphicspath{{Figs/}}

\title{Beamforming Design for RIS-Aided THz Wideband Communication Systems\\
}

\author{

Yihang $\text{Jiang}^{*}$, Ziqin $\text{Zhou}^{*}$, Xiaoyang $\text{Li}^{\dagger }$, and Yi $\text{Gong}^{*}$ \\
$^{*}$Southern University of Science and Technology, Shenzhen, China \\
$^{\dagger }$Shenzhen Research Institute of Big Data, Shenzhen, China\\
Email: 
gongy@sustech.edu.cn
}

\maketitle

\begin{abstract}
Benefiting from tens of GHz of bandwidth, terahertz (THz) communications has become a promising technology for future 6G networks. However, the conventional hybrid beamforming architecture based on frequency-independent phase-shifters is not able to cope with the beam split effect (BSE) in THz massive multiple-input multiple-output (MIMO) systems. Despite some work introducing the frequency-dependent phase shifts via the time delay network to mitigate the beam splitting in THz wideband communications, the corresponding issue in reconfigurable intelligent surface (RIS)-aided communications has not been well investigated. In this paper, the BSE in THz massive MIMO is quantified by analyzing the array gain loss. A new beamforming architecture has been proposed to mitigate this effect under RIS-aided communications scenarios. Simulations are performed to evaluate the effectiveness of the proposed system architecture in combating the array gain loss.
\end{abstract}


\section{Introduction}
%
Benefiting from ultra-high data rates, terahertz (THz) communication (0.1-10 THz) is considered as a key building block for future 6G networks \cite{song2011present}. 
However, due to the high carrier frequency, THz signals suffer from severe path loss, which is the main obstacle for THz communications to move from theory to practice \cite{han2015multi}.
To overcome the severe path loss of THz signals, directional beams 
with high array gains need to be generated by massive multiple-input multiple-output (MIMO) with hybrid beamforming (HB) architecture.

To achieve this goal, different phase shifts (PSs) provided by phase shifters (PSRs) array should be compensated at different antenna elements \cite{el2014spatially}.
It should be noted that the PSs to be compensated at a certain antenna element is frequency-dependent.
Since PSR is a frequency-independent device, the equiphase surfaces generated by the analog beamformer will be separated at different subcarriers.
Therefore, the beams generated by the analog beamformer will point to different physical directions surrounding the target physical direction at different subcarriers.
This effect is called ``\textbf{\textit{beam split}}'' in THz massive MIMO systems \cite{dai2022delay, gao2023integrated}.
In narrowband systems, the BSE can be ignored as the mainlobe of beam can still cover the target direction. 
As for wideband systems, splitting can seriously deteriorate the system performance. 
Some efforts have been made to mitigate beam splitting by introducing frequency-dependent PSs using time delay (TD) network \cite{ hashemi2008integrated, ghaderi2019integrated, dai2022delay}.
Specifically, each antenna element is controlled by time delayers (TDRs) in a one-to-one manner to realize wideband beamforming in \cite{hashemi2008integrated, ghaderi2019integrated}.
Dai et al. \cite{dai2022delay} analyzed the BSE in THz communication systems, and further proposed a practical architecture based on HB structure with a small number of TDRs. 
Arguably, the beam splitting issue in wideband communication systems has been greatly alleviated.

Additionally, the directional narrow beam is sensitive to the obstacles, which easily results in communication outages.
Note that reconfigurable intelligent surface (RIS) is an artificial meta-surface consisting of a large number of quasi-passive and low-cost reflective elements.
The phase of the incident electromagnetic waves reaching the reflective elements can be programmed to be electronically controlled so that the wireless environment can be configured in a favorable manner \cite{ wu2019towards}.
Due to its quasi-passive characteristics, RIS is free of self-interference and antenna noise amplification.
Therefore, one appealing and cost-effective solution to overcome the blockage issue is to deploy RISs.

In this paper, the BSE in RIS-aided THz wideband communications is studied. The fully connected HB architecture with TDRs and PSRs (TDRs-PSRs) is considered, and a new beamforming architecture is designed to combat the BSE in RIS-aided THz wideband communication systems.

\begin{figure}[htbp]
  \vspace{-0.4cm}
  \setlength{\abovecaptionskip}{0.cm}
  \centering
  \includegraphics[scale=0.4]{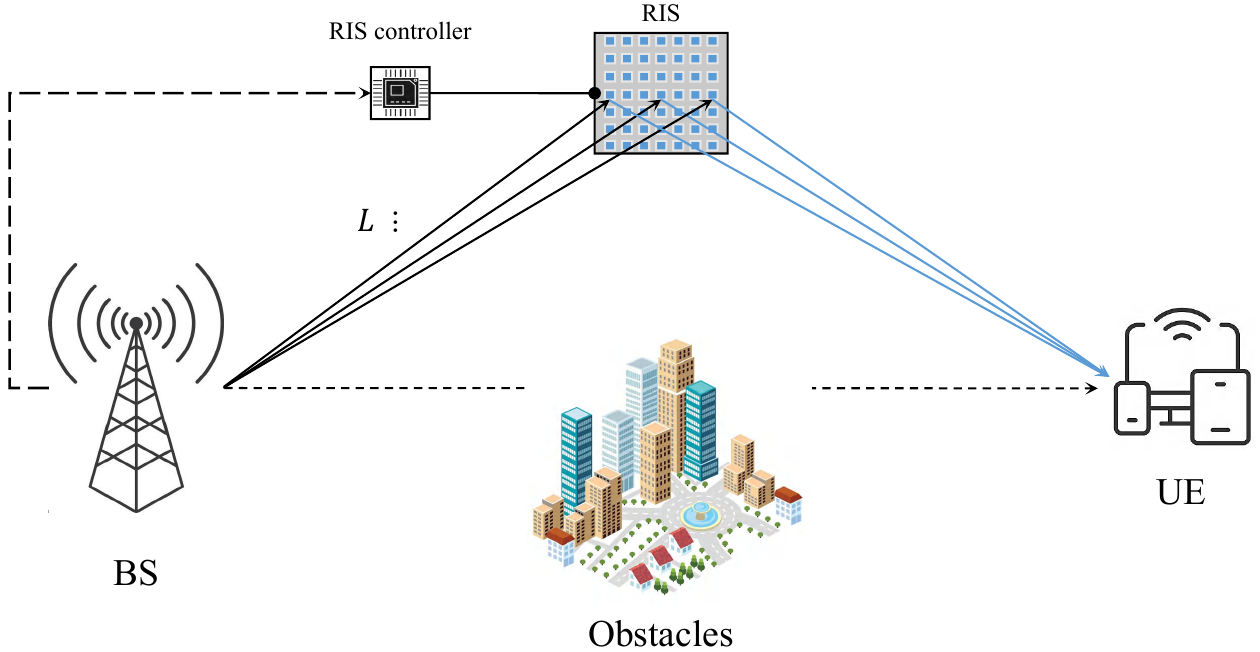}
  \caption{A RIS-aided downlink THz communication system.}
  \label{System Model}
  \vspace{-0.45cm}
\end{figure}

\begin{figure*}[htbp]
    \vspace{-0.6cm}
    \setlength{\abovecaptionskip}{0.cm}
    \centering
    \subfigure[The conventional HB architecture \cite{el2014spatially} \label{H}]
        {\includegraphics[width=0.32\linewidth]{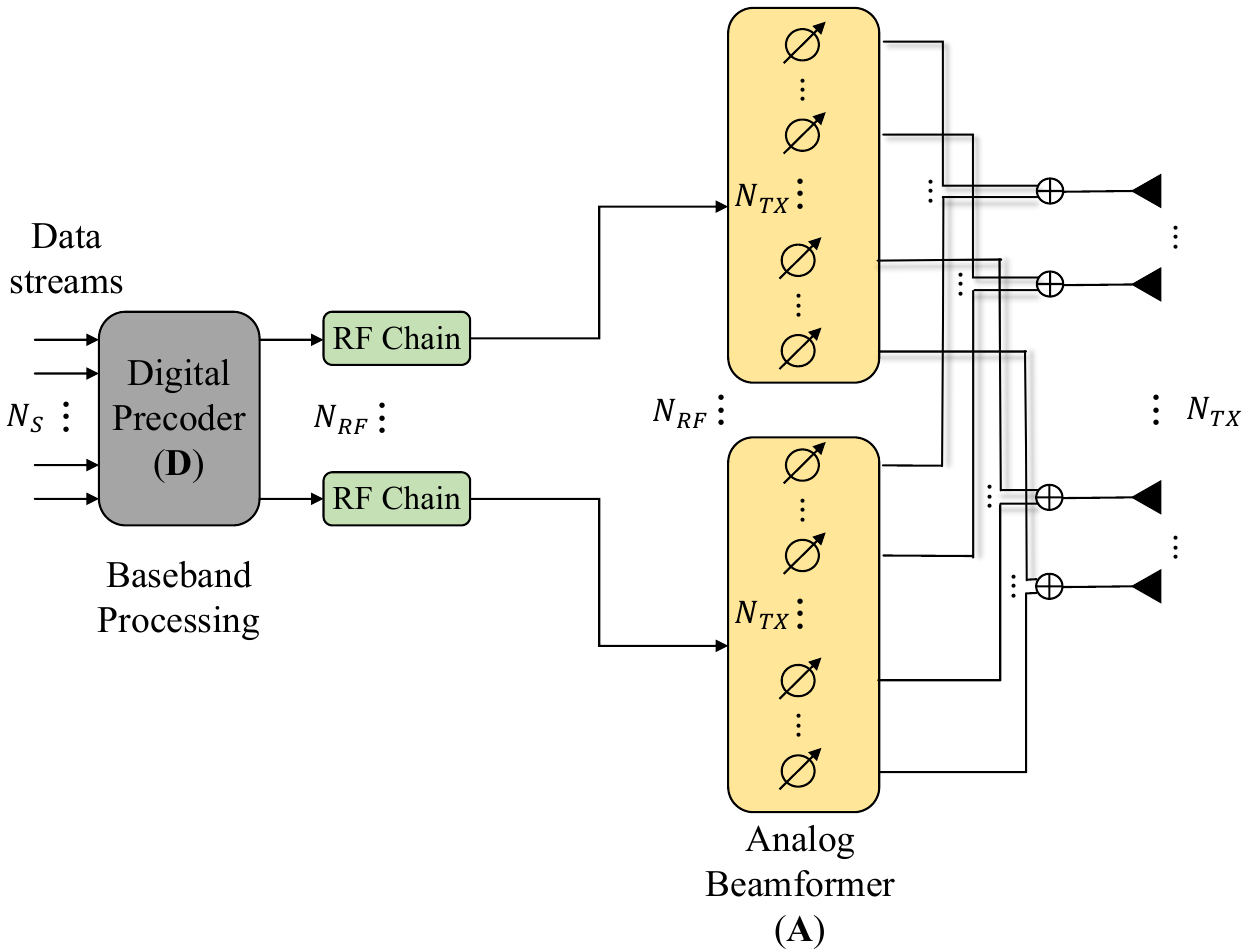}} 
    \qquad
        \subfigure[The fully connected TDRs-PSRs HB architecture \cite{dai2022delay} \label{HB2}]
        {\includegraphics[width=0.35\linewidth]{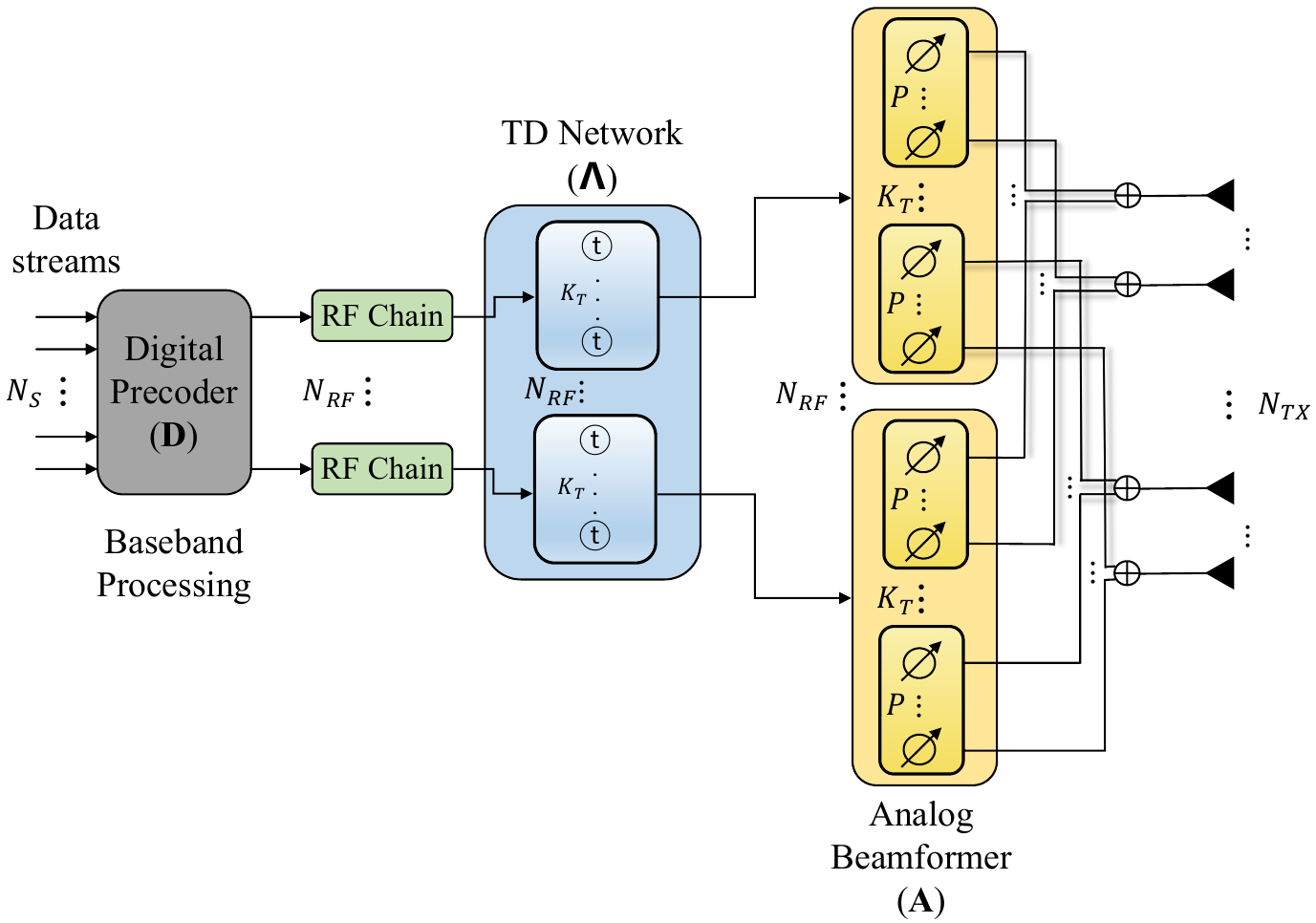}}  
    \caption{Comparison of HB architecture at the BS side.}
    \label{Base Station}
    \vspace{-0.5cm}
  \end{figure*}

\section{System Model}

\subsection{System Description}
The system architecture is indicated as Fig. \ref{System Model}.
We first consider a downlink RIS-aided wideband THz communication system where one BS equipped with $N_{TX}$ transmit antennas sends simultaneously $N_S$ data streams to an $N_{RX}$-antenna UE.
The BS employs the fully connected TDRs-PSRs HB architecture with $N_{RF}$ radio frequency (RF) chains.
As illustrated in Fig. \ref{Base Station}, compared with the conventional HB architecture, a TD network is introduced as a new precoding layer between the RF chains and the frequency-independent PSRs array.
Specifically, each RF chain is connected to a TDRs subarray consisting of $K_T$ TDRs, in which each TDR is connected to a PSRs subarray consisting of $P = \dfrac{N_{TX}}{K_T}$ frequency-independent PSRs in a sub-connected manner. 
The TD network can realize frequency-dependent PSs through TDRs, e.g., the PS $-2\pi ft$ can be achieved by the TD $t$ at the frequency $f$.
Thus, by utilizing the TD network, the fully connected TDRs-PSRs HB architecture converts the conventional phase-controlled beamformer into delay-phase jointly controlled beamformer, which can realize the frequency-dependent beamforming.

To solve the signal blockage problem, a RIS equipped with $F$ unit cells is deployed. The widely used orthogonal frequency division multiplexing (OFDM) with $M$ subcarriers is considered. We denote $f_c$ as the central frequency, $B$ as the bandwidth, and $f_m = f_c + \frac{B(2m-1-M)}{2M}$ as the frequency of the $m$-th subcarrier with $m\in  \left\{1, 2, \cdots, M\right\} $. The downlink received signal $\mathbf{y}_m \in \mathcal C^{N_{RX}\times 1}$ at the $m$-th subcarrier can be expressed as
\begin{equation}
  \label{eq1}
  \setlength{\abovedisplayskip}{3pt}
  \setlength{\belowdisplayskip}{3pt}  
  \mathbf{y}_m = \mathbf{H}_m\mathbf{\Psi } \mathbf{G}_m\mathbf{A}\mathbf{\Lambda}_m^{TD}\mathbf{D}_m\mathbf{s}_m + \mathbf{z}_m,
\end{equation}
where $\mathbf{s}_m \in \mathcal C^{N_S\times 1}$ is the transmitted signal at the $m$-th subcarrier with $\mathbb{E}(\mathbf{s}_m\mathbf{s}_m^H) = \dfrac{1}{N_S}\mathbf{I}_{N_S}$ and $\mathbf{D}_m\in \mathcal C^{N_{RF}\times N_S}$ denotes the frequency-dependent digital beamformer at the $m$-th subcarrier satisfying transmission power constraint $\sum_{m = 1}^{M}   \left\lVert \mathbf{A}\mathbf{\Lambda}_m^{TD}\mathbf{D}_m \right\rVert _F ^2 \leq  P_{total} $.
$\mathbf{\Lambda }_m^{TD}\in \mathcal C^{K_TN_{RF}\times N_{RF}} = blkdiag([e^{-j2\pi f_m\mathbf{t}_1}, \cdots, e^{-j2\pi f_m\mathbf{t}_n}, \cdots, e^{-j2\pi f_m\mathbf{t}_{N_{RF}}}])$ denotes the response matrix of the TD network,
where $\mathbf{t}_n \in \mathcal C^{K_T\times 1} = [t_{n,1}, \cdots, t_{n,K_T}]^T$ composes of the TDs realized by the $K_T$ TDRs connecting the $n$-th RF chain.
$\mathbf{A}\in \mathcal C^{N_{TX}\times K_TN_{RF}} = [\mathbf{A}_1, \cdots, \mathbf{A}_n, \cdots, \mathbf{A}_{N_{RF}}]$ with $\mathbf{A}_n \in \mathcal C^{N_{TX}\times K_T} = blkdiag([\mathbf{a}_{{PS}_{n,1}}, \cdots, \mathbf{a}_{{PS}_{n,k}}, \cdots, \mathbf{a}_{{PS}_{n,K_T}}])$ is the frequency-independent analog beamforming matrix identical over all $M$ subcarriers,
where $\mathbf{a}_{{PS}_{n,k}}\in \mathcal C^{P\times 1}$ composes of the frequency-independent PSs realized by the $k$-th PSRs subarray connecting the $n$-th RF chain.
$\mathbf{G}_m\in \mathcal C^{F\times N_{TX}}$ and $\mathbf{H}_m\in \mathcal C^{N_{RX}\times F}$ are the frequency domain channel responses of the BS-RIS and RIS-UE links at the $m$-th subcarrier, respectively.
$\mathbf{\Psi }\in \mathcal C^{F\times F}$ denotes the response matrix of the RIS and $\mathbf{z}_m\in \mathcal C^{N_{RX}\times 1}$ is the additive white Gaussian noise at the $m$-th subcarrier.

\subsection{Channel Model}
Channel estimation is not the main focus of this work and therefore we assume that all the channel state information (CSI) involved is perfectly known at the BS and RIS through the RIS controller as \cite{di2020hybrid}. Considering downlink transmissions, the wideband ray-based channel model \cite{dai2022delay} is adopted for THz communications. The frequency-domain channel response of the BS-RIS link can be expressed as
\begin{equation}
  \label{eq2}
  \setlength{\abovedisplayskip}{3pt}
  \setlength{\belowdisplayskip}{3pt}  
  \!\!\!\mathbf{G}_m = \sum\nolimits_{l_1=1}^{L_1} \alpha_{l_1}\mathbf{a}_{RIS}(\widetilde{\phi}_{l_1}^{RX}, \widetilde{\psi}_{l_1}^{RX}, f_m)\mathbf{a}_{BS}^H(\widetilde{\theta} _{l_1}^{TX}, f_m) ,
\end{equation}
where $\alpha_{l_1}$ represents the complex path gain of the $l_1$-th path and $L_1$ is the total number of the paths.
$\mathbf{a}_{BS}(\widetilde{\theta} _{l_1}^{TX}, f_m)$ is the transmit array steering vector of the BS for path $l_1$ at the subcarrier frequency $f_m$, where $\widetilde{\theta} _{l_1}^{TX}$ denotes the frequency-dependent spatial direction of angle of departure (AoD) for path $l_1$.
$\mathbf{a}_{RIS}(\widetilde{\phi} _{l_1}^{RX}, \widetilde{\psi} _{l_1}^{RX}, f_m)$ is the receive array steering vector of the RIS for path $l_1$ at the subcarrier frequency $f_m$, where $\widetilde{\phi}_{l_1}^{RX}$ and $\widetilde{\psi}_{l_1}^{RX}$ represent the frequency-dependent spatial directions of azimuthal and elevational angles of arrivals (AoAs) for path $l_1$, respectively.

In this paper, we consider the uniform linear array (ULA) and the uniform planar array (UPA) structure at the BS and RIS, respectively. Furthermore, $\sqrt{F}$ is assumed to be a positive integer, and the widely used square RIS is adopted in this paper. The relationship between the physical directions $(\widehat{\theta}_{l_1}^{TX}, \widehat{\phi}_{l_1}^{RX}, \widehat{\psi}_{l_1}^{RX})$ and the real directions $({\theta}_{l_1}^{TX}, {\phi}_{l_1}^{RX}, {\psi}_{l_1}^{RX})$ can be represented as $\widehat{\theta}_{l_1}^{TX} = \sin {\theta}_{l_1}^{TX}$, $\widehat{\phi}_{l_1}^{RX} = \sin {\phi}_{l_1}^{RX}$, and $\widehat{\psi}_{l_1}^{RX} = \sin {\psi}_{l_1}^{RX} \in \left[-1,1\right]$. The relationship between the spatial directions $(\widetilde{\theta} _{l_1}^{TX}, \widetilde{\phi}_{l_1}^{RX}, \widetilde{\psi}_{l_1}^{RX})$ and the physical directions $(\widehat{\theta}_{l_1}^{TX}$, $\widehat{\phi}_{l_1}^{RX}, \widehat{\psi}_{l_1}^{RX})$ can be represented as $\widetilde{\theta}_{l_1}^{TX} = 2d\frac{f_m}{c}\widehat{\theta}_{l_1}^{TX} = \frac{f_m}{f_c}\widehat{\theta}_{l_1}^{TX}$, $\widetilde{\phi}_{l_1}^{RX} = 2d\frac{f_m}{c}\widehat{\phi}_{l_1}^{RX} = \frac{f_m}{f_c}\widehat{\phi}_{l_1}^{RX}$ and $\widetilde{\psi}_{l_1}^{RX} = 2d\frac{f_m}{c}\widehat{\psi}_{l_1}^{RX} = \frac{f_m}{f_c}\widehat{\psi}_{l_1}^{RX}$, where $d$ is the antenna spacing set as $d = \frac{\lambda _c}{2} = \frac{c}{2f_c}$ with $\lambda _c$ being the wavelength at the central frequency $f_c$ and $c$ being the light speed. Consequently, the array steering vectors can be given as 
\begin{equation}
  \begin{aligned}
    \setlength{\abovedisplayskip}{3pt}
    \setlength{\belowdisplayskip}{3pt}  
  \label{eq3}
  \mathbf{a}_{BS}^H(\widetilde{\theta} _{l_1}^{TX}, f_m) = 1/\sqrt{N_{TX}} \mathbf{h}(\hat{\theta} _{l_1}^{TX}, f_m, N_{TX}),
  \end{aligned}
\end{equation}
and
\begin{align}
  \setlength{\abovedisplayskip}{3pt}
  \setlength{\belowdisplayskip}{3pt}  
  \label{eq4}
  &\mathbf{a}_{RIS}^H(\widetilde{\phi}_{l_1}^{RX}, \widetilde{\psi}_{l_1}^{RX}, f_m) = 1/\sqrt{F} \mathbf{h}(\phi_{l_1}^{RX}, \psi_{l_1}^{RX}, f_m, F),
\end{align}
where $\mathbf{h}(\hat{\theta} _{l_1}^{TX}\!\!\!,\! f_m\!,\! N_{TX}) \!=\! [1,\! e^{j\pi 2d\frac{f_m}{c}\widehat{\theta}_{l_1}^{TX}}\!\!\!,\! \cdots \!,\! e^{j\pi 2d\frac{f_m}{c} (N_{TX}-1)\widehat{\theta}_{l_1}^{TX}}\!]$, $\mathbf{h}(\phi_{l_1}^{RX}\!,\!  \psi_{l_1}^{RX}\!,\! f_m, F) \! =\!  [1, e^{j\pi 2d\frac{f_m}{c} (x\sin{\phi}_{l_1}^{RX} \sin{\psi}_{l_1}^{RX} + y\cos \psi_{l_1}^{RX})}, \\ \cdots, e^{j\pi 2d\frac{f_m}{c}((\sqrt{F}-1)\sin {\phi}_{l_1}^{RX} \sin{\psi}_{l_1}^{RX} + (\sqrt{F}-1)\cos \psi_{l_1}^{RX})} ]$ and $0 \!\leqslant\!  x, y \!\leqslant\! (\sqrt{F}\!-\!1)$.

Similarly, the frequency-domain channel response $\mathbf{H}_m$ between the RIS and UE can be presented as
\begin{equation}
  \setlength{\abovedisplayskip}{3pt}
  \setlength{\belowdisplayskip}{3pt}  
  \label{eq5}
 \!\!\! \mathbf{H}_m = \sum\nolimits_{l_2=1}^{L_2} \alpha_{l_2}\mathbf{a}_{UE}(\widetilde{\vartheta} _{l_2}^{RX}, f_m)\mathbf{a}_{RIS}^H(\widetilde{\phi} _{l_2}^{TX}, \widetilde{\psi}_{l_2}^{TX}, f_m) 
\end{equation}
Here, the signs involved in Eq. \eqref{eq5} are defined in the similar way as above.

Let $\mathcal F \triangleq \left\{1, 2, \cdots, F\right\}$ denote the set of all RIS unit cells and define the response matrix of the RIS as $\mathbf{\Psi} = diag(\Psi_1,  \cdots, \Psi_r, \cdots, \Psi_F)$, where $\Psi_r = \beta_r e^{j\varphi _r}$ with $\beta _r \in \left[0, 1\right]$ and $\varphi _r\in \left[0, 2\pi\right]$ are the amplitude reflection coefficient and the PS of the $r$-th unit cell respectively, which can be optimized by the RIS controller.
In this paper, we assume $\beta _r = 1$ for $ \forall r\in \mathcal{F} $ to maximize the signal reflection.

\section{Beamforming Design}
In this section, we analyze the BSE and design the corresponding beamforming for the above system.

\subsection{Analysis and Design of Combined Beamformers for the BS}
To better understand the BSE and its compensation mechanism, we take a detour and first analyze it in a conventional HB system.
As shown in the Fig. \ref{H}, in a conventional HB architecture, different PSs provided by PSRs could be compensated at different antenna elements so that all beams corresponding to different subcarriers are oriented in the same physical direction. 
Note that array steering vector $\mathbf{a}_{BS}^H(\widetilde{\theta} _{l_1}^{TX}, f_m) $ is frequency-dependent, which indicates that although the AoD $\widehat{\theta}_{l_1}^{TX}$ is fixed, the steering vector and the PS to be compensated will vary with frequency.
However, for PSRs-based analog beamforming, the PS of PSR is usually constant over the entire bandwidth because the PSR is a narrowband frequency-independent device.
Consequently, the analog beamforming direction corresponding to the maximum array gain should also varys with frequency.
Specifically, given an array steering vector $\mathbf{a}_{BS}^H(\widetilde{\theta} _{l_1}^{TX}, f_m)$, the normalized antennas array gain generated by the analog beamforming vector $\mathbf{\widehat{a} }_{PS}(\widehat{\theta} , f_c)$ in the physical direction $\widehat{\theta }_{l_1}^{TX} $ is a superposition of the signals radiated from all the antenna elements, which can be expressed as the \textit{array factor} (AF)
\begin{equation}
  \begin{aligned} 
    \setlength{\abovedisplayskip}{3pt}
    \setlength{\belowdisplayskip}{3pt}  
  \label{eq6}
\widehat{g}  _{AF}(\widehat{\theta} _{l_1}^{TX}, f_m, \widehat{\theta} , f_c) =  \left\lvert \mathbf{a}_{BS}^H(\widetilde{\theta} _{l_1}^{TX}, f_m) \mathbf{\widehat{a} }_{PS}(\widehat{\theta} , f_c) \right\rvert.
\end{aligned}
\end{equation}
Under the narrowband assumption, the normalized array gain $\widehat{g}  _{AF}(\widehat{\theta} _{l_1}^{TX}, f_m, \widehat{\theta} , f_c) \approx 1$ \footnote{In narrowband, $f_m \approx  f_c$. To achieve the maximum array gain, the analog beamforming vector should be aligned with the array steering vector, i.e., $\mathbf{\widehat{a} }_{PS}(\widehat{\theta}, f_c) = \mathbf{a}_{BS}(\widehat{\theta}^{TX}_{l_1}, f_c)$, and thus one have $\left\lvert \mathbf{a}_{BS}^H(\widetilde{\theta} _{l_1}^{TX}, f_m) \mathbf{a}_{BS}(\widetilde{\theta} _{l_1}^{TX}, f_m)\right\rvert = 1$.}. As for the wideband case, the normalized array gain achieves the maximum value 1 when the following condition is satisfied: $\widehat{\theta}^{TX}_{l_1} = \frac{f_c}{f_m}\widehat{\theta}$ \cite{gao2023integrated}. It is found that even though the intended beamforming direction is $\widehat{\theta} $, the actual matched physical direction $\widehat{\theta}^{TX}_{l_1}$ with maximum array gain will decrease as the frequency $f_m$ increases. Therefore, the conventional narrowband HB for mmWave cannot be directly applied for THz wideband communication.

Next, the BSE is analyzed in the fully connected TDRs-PSRs HB architecture. Considering that all RF chains are used to serve the same UE, then the analog beamforming matrix $\mathbf{A}_{n}$ corresponding to the $n$-th RF chain should be used to generate a directional beam in the physical direction $\widehat{\theta}_{n}^{TX} $ of the $n$-th path, which has been shown to be near-optimal \cite{dai2022delay, el2014spatially}. Since each RF chain has its own corresponding TD network and PSRs array, each individual RF chain is considered in the following analysis about array gain. By adopting the optimal analog beamforming vector $\mathbf{\widehat{a} }_{PS}(\widehat{\theta}, f_c) = \mathbf{a}_{BS}(\widehat{\theta}^{TX}_{l_1}, f_c)$ in the narrowband architecture here, the analog beamforming vector $\mathbf{a}^H_{{PS}_{n,k}}(\widehat{\theta}^{TX}_{n}, f_c)\in \mathcal C^{P\times 1}$ for the $n$-th RF chain controlled by the $k$-th subarray consisting of $P$ PSRs can be expressed as
\begin{equation}
    \begin{aligned}
  \label{eq7}
  \setlength{\abovedisplayskip}{3pt}
  \setlength{\belowdisplayskip}{3pt}  
    \mathbf{a}_{{PS}_{n,k}}^H(\widehat{\theta}^{TX}_{n} , f_c) = 1/\sqrt{N_{TX}}e^{j\pi (k-1)P\widehat{\theta} _{n}^{TX}} \mathbf{h}(\widehat{\theta} _{n}^{TX}, f_c, P) .
    \end{aligned}
\end{equation}
With the frequency-independent analog beamforming matrix $\mathbf{A}_{n}$ obtained from Eq. \eqref{eq7}, a beam aligned with the physical direction $\frac{f_c}{f_m}\mathbf{\widehat{\theta} }_{n}^{TX}$ at the $m$-th subcarrier frequency is generated.

Furthermore, the frequency-dependent TD network is introduced to rotate the physical direction, where the beam generated by $\mathbf{A}_{n}$ at the $m$-th subcarrier frequency with alignment from $\frac{f_c}{f_m}\widehat{\theta}_{n}^{TX}$ to $\widehat{\theta}_{n}^{TX}$.
The TDRs-PSRs (TP) array response $\mathbf{A}_m^{TP}\in \mathcal C^{N_{TX}\times N_{RF}}$ can be presented as
\begin{equation}
  \begin{aligned} 
  \label{eq8}
  \setlength{\abovedisplayskip}{3pt}
  \setlength{\belowdisplayskip}{3pt}  
  \!\!\!\!\mathbf{A}_m^{TP} \!=\! \mathbf{A}\mathbf{\Lambda }_m^{TD} 
  \!=\! [\mathbf{a}_{{TP}_{1,m}}, \cdots, \mathbf{a}_{{TP}_{n,m}}, \cdots, \mathbf{a}_{{TP}_{N_{RF},m}}],
\end{aligned}
\end{equation}
where $\mathbf{a}_{{TP}_{n,m}} \in \mathcal C^{N_{TX}\times 1} = blkdiag([\mathbf{a}_{{PS}_{n,1}} , \cdots , \mathbf{a}_{{PS}_{n,K_T}}])\\ \cdot \mathbf{T}_{n,m} = [\mathbf{a}_{{TP}_{n,m[1,1]}}, \cdots, \mathbf{a}_{{TP}_{n,m[k,p]}}, \cdots, \mathbf{a}_{{TP}_{n,m[K_T,P]}}]^T$ with $\mathbf{T}_{n,m}^H \in \mathcal C^{1\times K_T} = [e^{j2\pi f_mt_{n,1}}, \cdots, e^{j2\pi f_mt_{n,K_T}}]$ as the frequency-dependent PSs vector for the $n$-th RF chain,
and the element $\mathbf{a}_{{TP}_{n,m[k,p]}}$ is generated by the $p$-th PSR connected to the $k$-th TDR satisfying $\mathbf{a}_{{TP}_{n,m[k,p]}} = \frac{1}{\sqrt{N_{TX}}}e^{-j2\pi \left\{[(k-1)P+p-1]\frac{1}{2}\sin \theta ^{TX}_{n}+f_mt_{n,k}\right\} }$ for $\forall k = [1, \cdots, K_T], p = [1, \cdots, P]$.
Then the array gain after the TP array can be rewritten as 
\begin{align} 
  \label{eq9}
  \setlength{\abovedisplayskip}{3pt}
  \setlength{\belowdisplayskip}{3pt}  
  &\!\! g_{AF}(\widetilde{\theta }_{n}^{TX}, f_m) = \left\lvert \sum\nolimits_{k = 1}^{K_T} \sum\nolimits_{p = 1}^{P} \mathbf{a}_{{TP}_{n,m[k,p]}}\mathbf{a}^H_{{BS}_{[k,p]}}   \right\rvert \\
 &\!\!\!\! = \left\lvert \sum\nolimits_{k = 1}^{K_T} \sum\nolimits_{p = 1}^{P} \frac{1}{N_{TX}}e^{j\pi \left\{[(k-1)P+p-1](\frac{f_m}{f_c}-1) \widehat{\theta}  _{n}^{TX} - 2f_mt_{n,k}\right\}  }   \right\rvert . \nonumber
\end{align}
It can be observed that the normalized array gain in Eq. \eqref{eq9} can achieve the maximum value 1 when the following condition is satisfied, i.e.,
\begin{equation}
  \begin{aligned}
  \label{eq10}
  \setlength{\abovedisplayskip}{3pt}
  \setlength{\belowdisplayskip}{3pt}  
  t_{n,k} &=   [(k-1)P+p-1](f_m/f_c - 1)\widehat{\theta} _{n}^{TX}/(2f_m) .
\end{aligned}
\end{equation}
As $t_{n,k}$ depends on specific $p$, the optimal TDR should serve the antenna element in an one-to-one manner. However, due to the economic issues, the one-to-many manner is adopted at the BS, and the TD can be further obtained as
\begin{equation}
  \begin{aligned}
  \label{eq11}
  \setlength{\abovedisplayskip}{3pt}
  \setlength{\belowdisplayskip}{3pt}  
  t_{n,k} = (f_m/f_c - 1)(k-1)P\widehat{\theta }_{n}^{TX}/(2f_m) ,
\end{aligned}
\end{equation}
where the TDR focuses on perfectly compensating the PS of the first PSR in each PSRs subarray by setting $p = 1$. In other words, after the beam alignment of the sub-connected TP array, there is still a slight loss of array gain for different subcarriers except for the central subcarrier. Specifically, substituting Eq. \eqref{eq11} into $\mathbf{a}_{{TP}_{n,m[k,p]}}$, the retained item after TP array at the $m$-th subcarrier should be
\begin{align}
  \label{eq12}
  \setlength{\abovedisplayskip}{3pt}
  \setlength{\belowdisplayskip}{3pt}  
  \zeta _m &= \sum\nolimits_{k = 1}^{K_T} \sum\nolimits_{p = 1}^{P} \frac{1}{N_{TX}} e^{j \pi {[(k-1)P+p-1](\frac{f_m}{f_c}-1) \widehat{\theta}  _{n}^{TX} - 2f_mt_{n,k}} } \nonumber \\
  &\overset{\text{(a)} }{= } e^{j\frac{(P-1)\pi}{2}(\frac{f_m}{f_c}-1) \widehat{\theta} _{n}^{TX}}\Xi _{P}\left[(f_m/f_c - 1) \widehat{\theta} _{n}^{TX}\right]/P ,
\end{align}
where (a) comes from the equation $\sum_{t = 0}^{N-1} e^{jt\pi a} = \frac{\sin \frac{N\pi}{2}a}{\sin \frac{\pi}{2}a}e^{j\frac{(N-1)\pi}{2}a} $, and $\Xi _N(x) = \frac{\sin \frac{N\pi}{2}x}{\sin \frac{\pi}{2}x}$ with $\left\lvert \Xi _N(0)\right\rvert  = N$ is the Dirichlet sinc function \cite{sayeed2002deconstructing}.
Clearly, for the non-central subcarriers, the introduced array gain loss at the $m$-th subcarrier in the physical direction of RIS should be 
\begin{equation}
  \label{eq13}
  \setlength{\abovedisplayskip}{3pt}
  \setlength{\belowdisplayskip}{3pt}  
g_{loss_m}^{BS} = 1 - \left\lvert \Xi _{P} \left[(f_m/f_c - 1) \widehat{\theta}  _{n}^{TX}\right] \right\rvert/P .
\end{equation}

It can be observed that the array gain loss in a given physical direction depends on the carrier frequency as well as bandwidth and the number of TDRs.

\subsection{Analysis and Design of Response Matrix for the RIS }
The receive raw signal at the RIS can be presented as 
\begin{equation}
  \begin{aligned}
  \label{eq14}
  \setlength{\abovedisplayskip}{3pt}
  \setlength{\belowdisplayskip}{3pt}  
  \mathbf{\widehat{y} } _m 
    =  \mathbf{G}_m\mathbf{A}_m^{TP}\mathbf{D}_m\mathbf{s}_m + \mathbf{z}_m = \mathbf{\widehat{G} }_m\mathbf{D}_m\mathbf{s}_m + \mathbf{z}_m ,
  \end{aligned}
\end{equation}
where $\widehat{\mathbf{G}}_m \in \mathcal C^{F\times N_{RF}} = [\mathbf{g}_{1,m}, \cdots, \mathbf{g}_{n,m}, \cdots, \mathbf{g}_{N_{RF},m}]$ denotes the equivalent channel response of BS-RIS link at the $m$-th subcarrier frequency with $ \mathbf{g}_{n,m} \in \mathcal C^{F\times 1} = \sum_{l_1=1}^{L_1} \alpha_{l_1}\mathbf{a}_{RIS}(\widetilde{\phi}_{l_1}^{RX} \!\!,\! \widetilde{\psi}_{l_1}^{RX} \!\!,\! f_m)  \mathbf{a}_{BS}^H(\widetilde{\theta} _{l_1}^{TX} \!\!,\! f_m) \mathbf{a}_{{TP}_{n, m}} $ denotes the equivalent channel vector for the $n$-th RF chain. 
Further, considering the BS and RIS as a whole, $\widehat{\mathbf{A}}_m \triangleq  \mathbf{\Psi }\mathbf{\widehat{G} }_m = \mathbf{\Psi }\left[\mathbf{g }_{1, m}, \cdots, \mathbf{g }_{n, m}, \cdots, \mathbf{g}_{N_{RF}, m}\right] = \left[\widehat{\mathbf{A}}_{1,m}, \cdots, \widehat{\mathbf{A}}_{n,m}, \cdots,  \widehat{\mathbf{A}}_{N_{RF},m}\right] $ denotes the frequency-dependent equivalent RF response at the $m$-th subcarrier frequency.

Similarly, focusing on the $n$-th RF chain, the equivalent normalized array gain at the $m$-th subcarrier frequency in physical direction $(\widehat{\phi}  _{l_2}^{TX}, \widehat{\psi}  _{l_2}^{TX})$ can be presented as
\begin{equation}
\label{eq15}
\setlength{\abovedisplayskip}{3pt}
\setlength{\belowdisplayskip}{3pt}  
   \!\!\! g_{AF}(\widehat{\phi}_{l_2}^{TX}, \widehat{\psi}_{l_2}^{TX}, f_m) = \left\lvert \mathbf{a}_{RIS}^H(\widetilde{\phi}_{l_2}^{TX}, \widetilde{\psi}_{l_2}^{TX}, f_m)\widehat{\mathbf{A}}_{n,m}\right\rvert .
\end{equation}
The normalized array gain achieves the maximum value 1 when $\widehat{\mathbf{A}}_{n,m} = \mathbf{a}_{RIS}(\widetilde{\phi}_{l_2}^{TX}, \widetilde{\psi}_{l_2}^{TX}, f_m)$ is satisfied.
On the other hand, to maximize the mutual information achieved by Gaussian signaling over the RIS-UE channel, for $\forall l_2 \in \left\{1, \cdots, L_2 \right\} $, $N_{RF}$ of the vectors $a_{RIS}(\widetilde{\phi}_{l_2}^{TX}, \widetilde{\psi}_{l_2}^{TX}, f_m)$ should be used as columns of $\widehat{\mathbf{A}}_m$ \cite{el2014spatially}.
In addition, define the ordered singular value decomposition (SVD) of the RIS-UE channel at the $m$-th subcarrier frequency as $\mathbf{H}_m = \mathbf{U}\mathbf{\Sigma }\mathbf{V}^H$.
The first $rank(\mathbf{H}_m)$ right singular vectors arranged in decreasing order in $\mathbf{V}$ can be linearly represented by $\sum_{l_2 = 1}^{L_2} \mathbf{a}_{RIS}(\widetilde{\phi}_{l_2}^{TX}, \widetilde{\psi}_{l_2}^{TX}, f_m) $, where $\mathbf{a}_{RIS}(\widetilde{\phi}_{l_2}^{TX}, \widetilde{\psi}_{l_2}^{TX}, f_m)$ with path gain arranged in decreasing order, is the main component of the $l_2$-th singular vector for $\forall l_2 = \left\{1, 2, \cdots, L_2 \right\} $  \cite{jiang2022design} \footnote{In this paper, we assume that $L_2 \leq min(N_{RX}, F)$. That is reasonable, considering that THz communication relies heavily on the line of sight (LoS) path \cite{tan2021wideband}.}.
Therefore, the $l_2$-th column vector $\widehat{\mathbf{A}}_{l_2,m}$ should be designed to generate a directional beam towards the $l_2$-th path's physical direction $(\widehat{\phi }^{TX}_{l_2}, \widehat{\psi }^{TX}_{l_2} )$ for $\forall m = \left\{1, 2, \cdots, M\right\} $. 

The problem is converted to make $\widehat{\mathbf{A}}_{l_2,m}$ sufficiently “close” to the optimal precoder $\mathbf{a}_{RIS}(\widetilde{\phi}_{l_2}^{TX}, \widetilde{\psi}_{l_2}^{TX}, f_m)$, which can be equivalently stated as 
\begin{align}
  \label{eq16}
  \setlength{\abovedisplayskip}{3pt}
  \setlength{\belowdisplayskip}{3pt}  
  \mathbf{P1} : \mathbf{\Psi } &= \arg\min\limits_{\mathbf{\Psi }} \left\lVert \mathbf{a}_{RIS}(\widetilde{\phi}_{l_2}^{TX}, \widetilde{\psi}_{l_2}^{TX}, f_m) - \widehat{\mathbf{A}}_{l_2,m}\right\rVert _F^2 \\
  \text{s.t.}~~ 
  \mathbf{\Psi } &= diag(\Psi_1, \cdots \!,\! \Psi_r, \cdots\!,\! \Psi_F), \Psi _r = e^{j\varphi_r}\!,\! \varphi_r\in \left[0, 2\pi\right] \nonumber.
\end{align}

However, the RIS can only change the phase like the PSR, and it is hard to meet the alignment of each subcarrier frequency.
Let the beam of the central carrier towards the physical direction of the UE, an exact solution $\mathbf{\Psi }_0$ to $\mathbf{(P1)}$ can be obtained. Assuming that the BSE at the BS is perfectly mitigated by setting $\mathbf{a}_{{TP}_{l_2, m}} = \mathbf{a}_{BS}(\widetilde{\theta} _{l_2}^{TX} , f_m)$, $\widehat{\mathbf{A}}_{l_2,m}$ can be approximately represented as
\begin{align}
\label{eq17}
\setlength{\abovedisplayskip}{3pt}
\setlength{\belowdisplayskip}{3pt}  
&\widehat{\mathbf{A}}_{l_2,m} = \mathbf{\Psi }\mathbf{g}_{l_2,m} \approx \nonumber \\
&\! \mathbf{\Psi } \! \sum\nolimits_{l_1=1}^{L_1} \! \alpha_{l_1}\mathbf{a}_{RIS}(\widetilde{\phi}_{l_1}^{RX} \!,\! \widetilde{\psi}_{l_1}^{RX} \!,\! f_m) \mathbf{a}_{BS}^H(\widetilde{\theta} _{l_1}^{TX} \!,\! f_m) \mathbf{a}_{BS}(\widetilde{\theta }_{l_2}^{TX} \!,\! f_m) \nonumber \\
&= \frac{\mathbf{\Psi }}{N_{TX}^2 }\sum\nolimits_{l_1 = 1}^{L_1} \! a_{l_1} \Xi _{N_{TX}}(\widetilde{\theta }_{l_1}^{TX}-\widetilde{\theta }_{l_2}^{TX})e^{-j\frac{N_{TX}-1}{2}\pi(\widetilde{\theta }_{l_1}^{TX}-\widetilde{\theta }_{l_2}^{TX})} \nonumber \\
&\cdot  \mathbf{a}_{RIS}(\widetilde{\phi}_{l_1}^{RX}, \widetilde{\psi}_{l_1}^{RX}, f_m) .
\end{align}

From Eq. \eqref{eq17}, the element $\widehat{\mathbf{A}}_{l_2,c[x,y]}$ corresponding to the $x$-th row and the $y$-th column of the RIS with the subcarrier frequency $f_c$ can be presented as 
\begin{align}
\label{eq18}
\setlength{\abovedisplayskip}{3pt}
\setlength{\belowdisplayskip}{3pt}  
& \widehat{\mathbf{A}}_{l_2,c[x,y]} = \frac{\mathbf{\Psi }_{[x,y]}}{N^2_{TX}\sqrt{F}}\sum\nolimits_{l_1=1}^{L_1} \alpha_{l_1} \Xi _{N_{TX}}(\widetilde{\theta }_{l_1}^{TX}-\widetilde{\theta }_{l_2}^{TX}) \nonumber\\
&\cdot  e^{-j\pi[\frac{N_{TX}-1}{2}(\widehat{\theta}_{l_1}^{TX} - \widehat{\theta}_{l_2}^{TX}) + (x\widehat{\phi}_{l_1}^{RX}\widehat{\psi}_{l_1}^{RX} + y\cos \psi_{l_1}^{RX})]} .
\end{align}

Substituting Eq. \eqref{eq18} into $\widehat{\mathbf{A}}_{l_2,c} = \mathbf{a}_{RIS}(\widetilde{\phi}_{l_2}^{TX}, \widetilde{\psi}_{l_2}^{TX}, f_c)$, then the exact solution $\mathbf{\Psi }_{0[x,y]}$ can be obtained as Eq. \eqref{eq19}.
\begin{figure*}
\vspace{-0.6cm}
\setlength{\abovedisplayskip}{3pt}
\setlength{\belowdisplayskip}{3pt}
    \begin{equation}
  \label{eq19}
  \mathbf{\Psi }_{0[x,y]} = \frac{N^2_{TX}Fe^{-j\pi (x \widehat{\phi} _{l_2}^{TX} \widehat{\psi} _{l_2}^{TX} + y\cos \psi_{l_2}^{TX})}}{\sum_{l_1=1}^{L_1} \alpha_{l_1} \Xi _{N_{TX}}(\widetilde{\theta }_{l_1}^{TX}-\widetilde{\theta }_{l_2}^{TX})e^{-j\pi[\frac{N_{TX}-1}{2}(\widehat{\theta} _{l_1}^{TX} - \widehat{\theta} _{l_2}^{TX}) + (x \widehat{\phi} _{l_1}^{RX} \widehat{\psi} _{l_1}^{RX} + y\cos \psi_{l_1}^{RX})]}} .
\end{equation}
\hrulefill
\vspace{-0.5cm}
\end{figure*}
It can be observed that the response matrix $\mathbf{\Psi }_0$ depends on the $l_2$-th path component. As each RF-chain is designed to form a beam towards one path component, each RF chain should have a different $\mathbf{\Psi }_0$. Ideally, each RF chain have a dedicated RIS to serve it in an one-to-one manner. The $l_2$-th RF chain should be reflected by its dedicated RIS with an elaborate response matrix in order to direct the beam at the frequency $f_c$ towards the physical direction of the $l_2$-th path component.

\section{Simulation and Analysis}

\begin{table}[htbp]
  \setlength{\abovedisplayskip}{3pt}
  \setlength{\belowdisplayskip}{3pt}  
  \caption{System Parameters for Simulations}
  \vspace{-0.2cm}
  \centering
  \resizebox{.9\columnwidth}{!}{
  \begin{tabular}{ c  c  c } 
    \toprule
      \textbf{Symbol} & \textbf{Quantity} & \textbf{Parameter Description}\\
    \midrule
      $N_{TX}$ & 256 & Number of the transmit antennas\\
      $N_{RX}$ & 64 &  Number of the receive antennas\\ 
      $B$ & 30 GHz & Bandwidth\\ 
      $f_c$ & 300 GHz & Central subcarrier frequency \\ 
      $K_T$ & 16, 32 & Number of TD elements \\
      $F$ & 64 $(8\times 8)$ & Number of RIS elements \\
      $(\widehat{\theta}^{TX})$ & $(0.5)$ & Physical direction of AoD at the BS \\ 
      $(\widehat{\phi}^{RX}, \widehat{\psi}^{RX})$ & $(0.4, 0.5)$ & Physical directions of AoAs at the RIS \\ 
      $(\widehat{\phi}^{TX}, \widehat{\psi}^{TX})$ & $(0.5, \frac{\sqrt{3}}{2})$ & Physical directions of AoDs at the RIS \\ 
    \bottomrule
  \end{tabular}}
  \label{tab1}
\end{table}

\begin{figure}[htbp]
\vspace{-0.2cm}
\setlength{\abovecaptionskip}{0.cm}
\centering
\subfigure[Conventional HB architecture  \label{fig1}]
    {\includegraphics[width=0.47\linewidth]{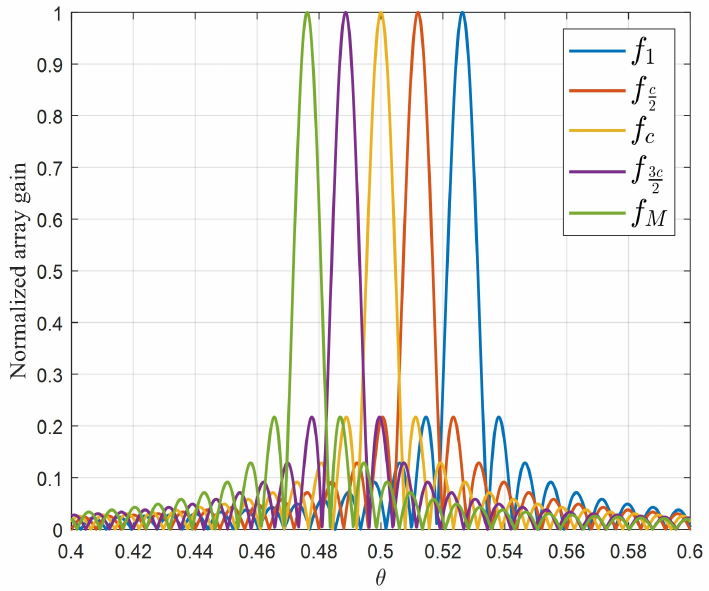}}
\qquad
\subfigure[HB with TD network  \label{fig2}]
    {\includegraphics[width=0.435\linewidth]{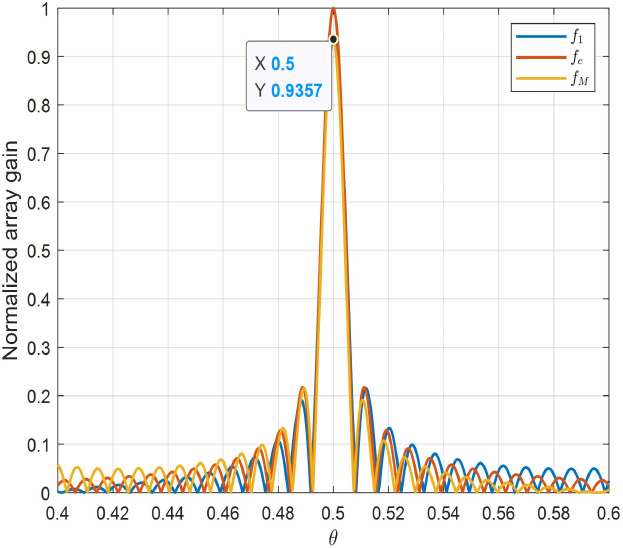}}  
\caption{Comparison of array gain for two HB architectures.}
\vspace{-0.4cm}
\label{Fig3}
\end{figure}

\begin{figure}[htbp]
  \setlength{\abovecaptionskip}{0.cm}
    \centering
    \includegraphics[scale = 0.4]{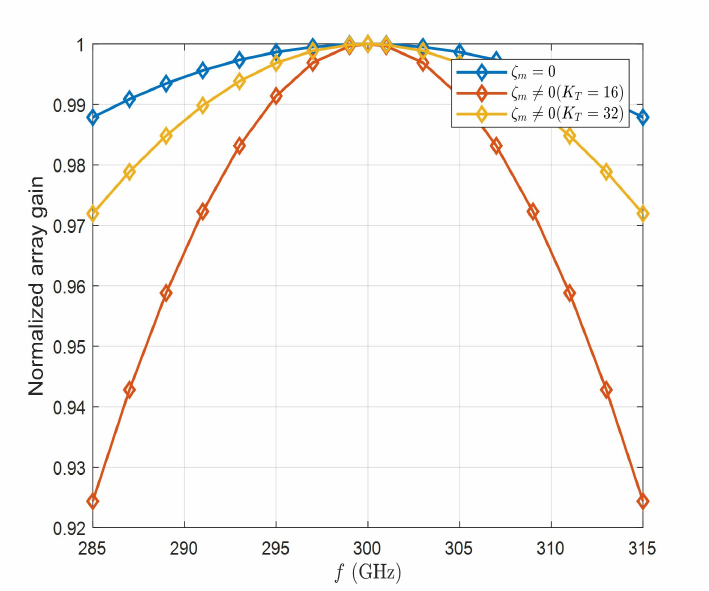}
    \caption{Comparison of array gains at the RIS side.}
    \label{Fig4}
    \vspace{-0.4cm}
  \end{figure}

The parameter settings are shown in TABLE \ref{tab1}.    
To better illustrate the BSE, we compare the normalized array gain between conventional HB architecture and HB with TD network in Fig. \ref{Fig3}.
Fig. \ref{fig1} corresponds to the results of Eq. \eqref{eq7}, and it is clear that the conventional HB structure fails to resolve the BSE at the THz frequency band.
Severe beam splitting occurs at different subcarrier frequencies, and array gain is difficult to accumulate effectively in the physical direction of the UE.
Fig. \ref{fig2} corresponds to the results of Eq. \eqref{eq11}.
Apparently, although there is still some array gain loss for non-central carriers, the introduction of TD networks has greatly mitigated the BSE.
The array gain at the RIS side are shown in Fig. \ref{Fig4}. 
With our proposed beamforming scheme, the array gain loss over the entire frequency band is quit small, and the array gain can be further improved by adding a small amount of TDRs.  
Therefore, the BSE under the blockage scenario is well tackled.

\section{Concluding Remarks}
In this paper, we present an architecture for jointly designing beamforming of BS and RIS in the signal blockage case.
The simulations confirm that the beam splitting has been well alleviated. 
Moving forward, it would be of interest to further consider the achievable rate with the proposed architecture.
Given the numerous parameters involved, the optimization problem of the achievable rate will be very complex, which will be fully investigated in our subsequent work.

\section*{Acknowledgment}
This work was supported in part by National Natural Science Foundation of China under Grants 62071212 and 62101235, Shenzhen Science and Technology Program under Grant JCYJ20220530113017039, and the internal Project Fund from Shenzhen Research Institute of Big Data under Grants J00120230001 and J00220230004.

\normalem
\bibliography{refs}
\bibliographystyle{ieeetr}

\end{document}